\documentclass[journal abbreviation]{copernicus}

\newcommand{\be}{\begin{equation}}
\newcommand{\ee}{\end{equation}}

\newcommand{\eqs}[2]{Eqs. (\ref{#1}) \& (\ref{#2})}

\newcommand{\eq}[1]{Eq. (\ref{#1})} 
 
\newcommand{\Fig}[1]{Fig. (\ref{#1})} 
 \newcommand{\eqa}{\begin{eqnarray}}
\newcommand{\eeq}{\end{eqnarray}}

\begin{document}

\title{Spectral Properties of Electromagnetic Turbulence in Plasmas}

\author[1]{Dastgeer Shaikh}
\author[2]{P. K. Shukla}

\affil[1]{Department of Physics and Center for Space Plasma and Aeronomy Research (CSPAR),
The University of Alabama in Huntsville, Huntsville, AL-35899. USA}
\affil[2]{Institut f\"ur Theoretische Physik IV, Fakult\"at f\"ur Physik und
Astronomie, Ruhr-Universit\"at Bochum, D-44780 Bochum, Germany, 
and SUPA Department of Physics, The University of Strathclyde, Glasgow, 
Scotland, United Kingdom}


\runningtitle{Electromagnetic Turbulence}

\runningauthor{Shaikh \& Shukla}

\correspondence{Dastgeer Shaikh \\(dastgeer.shaikh@uah.edu)\\
P. K. Shukla (ps@tp4.rub.de)}

\received{11 December 2009}
\pubdiscuss{} 
\revised{}
\accepted{}
\published{}


\firstpage{1}

\maketitle

\begin{abstract}
We report on the nonlinear turbulent processes associated with
electromagnetic waves in plasmas. We focus on low-frequency (in
comparison with the electron gyrofrequency) nonlinearly interacting
electron whistlers and nonlinearly interacting Hall-magnetohydrodynamic 
(H-MHD) fluctuations in a magnetized plasma.
Nonlinear whistler mode turbulence study in a magnetized plasma
involves incompressible electrons and immobile ions. Two-dimensional
turbulent interactions and subsequent energy cascades are critically
influenced by the electron whisters that behave distinctly for 
scales smaller and larger than the electron skin depth. It is found
that in whistler mode turbulence there results a dual cascade
primarily due to the forward spectral migration of energy that
coexists with a backward spectral transfer of mean squared magnetic
potential. Finally, inclusion of the ion dynamics, resulting from a
two fluid description of the H-MHD plasma, leads to several
interesting results that are typically observed in the solar wind
plasma.  Particularly in the solar wind, the high-time-resolution
databases identify a spectral break at the end of the MHD inertial range
spectrum that corresponds to a high-frequency regime. In the latter,
turbulent cascades cannot be explained by the usual MHD model and a
finite frequency effect (in comparison with the ion gyrofrequency) arising 
from the ion inertia is essentially included to discern the dynamics of the 
smaller length scales (in comparison with the ion skin depth). This leads 
to a nonlinear H-MHD model, which is presented in this paper.  With the help of 
our 3D H-MHD code, we find that the characteristic turbulent interactions in 
the high-frequency regime evolve typically on kinetic-Alfv\'en time-scales. The turbulent
fluctuation associated with kinetic-Alfv\'en interactions are
compressive and anisotropic and possess equipartition of the kinetic
and magnetic energies.
\end{abstract}


\section{Introduction}
Many laboratory and space plasmas contain multi-scale electromagnetic
fluctuations. The latter include the electron whistlers and H-MHD
fluctuations in a uniform magnetoplasma.  While the wavelengths of the
electron whistlers could be comparable with the electron skin depth,
the H-MHD fluctuations could have differential wavelengths when
compared with the ion skin depth and the ion-sound gyroradius.  In
whistler mode turbulence, the electrons are incompressible and the
ions form the neutralizing background in the plasma. On the other
hand, in the H-MHD turbulence the compressibility of the electrons and
ions cannot be ignored.

In uniform magnetized plasmas, one can have different types of
electromagnetic waves. The latter include the low-frequency (in
comparison with the electron gyrofrequency) electron whistlers as well
as linearly coupled fast and slow magnetosonic and kinetic-Alfv\'en
waves. The electron whistlers are described by the electron
magnetohydrodynamics (EMHD) equations \cite{kingsep}, and the role of
whistlers in the EMHD turbulence is an unresolved issue
\cite{dastgeer1,dastgeer2,dastgeer3,dastgeer4,dastgeer5,dastgeer6,dastgeer2009}. Notably,
EMHD in two dimensions exhibits dual cascade phenomena
\cite{biskamp96} similar to those of 2D Navier-Stokes \citep{krai,fri}
and MHD dynamics \citep{fyfe,biskamp93}. However, because of the
complexity of nonlinear interactions in EMHD turbulence, mode coupling
interactions are relatively more complex to comprehend.  Like 2D
Navier-Stokes \citep{krai,fri} and MHD \citep{fyfe,biskamp93} systems,
2D EMHD also exhibits dual cascade processes \citep{biskamp96}.  The
dual cascade phenomena, i.e. both the forward and inverse cascades, is
however difficult to resolve numerically in a single energy spectrum
because of a lack of spectral resolution in the inertial ranges.
Therefore there exists considerably less numerical work on this issue
as compared to various two-dimensional (decaying) turbulence
systems. Numerical simulations of inverse cascade phenomenon, within
the context of magnetohydrodynamics (MHD) turbulence, have previously
been performed at a modest resolution of up to $1024^2$ Fourier modes
\citep{biskamp93}.

The dynamics of linearly coupled magnetic field-aligned Alfv\'en wave and ´
obliquely propagating fast and slow magnetosonic waves and kinetic Alfv\'en waves 
are governed by the H-MHD equations.  The solar
wind plasma wave spectra corresponding to whistlers or high-frequency
(in comparison with the ion gyrofrequency) kinetic-Alfv\'en wave
possess a spectral break, the origin of which is not yet fully
understood \cite{danskat,goldstein1,leamon,goldstein2,bale05,servidio08}. 
There are, however, some preliminary results available in the literature
\citet{dastgeer7,dastgeer8}.

In this paper, we present a survey of the spectral properties of the
low-frequency multi-scale electromagnetic turbulence in plasmas.  In
section 2, we begin by describing the whistler wave model based on two
dimensional electron magnetohydrodynamic equations. Linear properties
of whistler waves are described in this section. Section 3 contains
nonlinear simulation results describing the wave spectra in small and
large scale (in comparison with the electron skin depth) regimes.  
Section 4 describes the Hall- MHD counter part.
The summary and conclusions are contained in section 5.

\section{The Whistler Mode Turbulence}

In an incompressible electron-MHD (E-MHD) plasma, collective
oscillation of electrons exhibit electron whistlers. The electron
Whistlers are widely observed in many space and laboratory
plasmas. For instance, they are believed to be generated in the
Earth's ionospheric region by lightning discharges and proceed in the
direction of Earth's dipole magnetic field \citep{helli}. They have
also been recently detected in the Earth's radiation belt by the
STEREO S/WAVES instrument \cite{cattell}.  Moreover, there are
observations of Venus' ionosphere that reveal strong, circularly
polarized, electromagnetic waves with frequencies near 100 Hz. The
waves appear as bursts of radiation lasting 0.25 to 0.5s, and have the
expected properties of whistler-mode signals generated by lightning
discharges in Venus' clouds \cite{Russell}. These waves are also
reported near the Earth's magnetopause \cite{Stenberg} and the Cluster
spacecraft encountered them during the process of magnetic
reconnection in the Earth's magnetotail region \cite{wei}.  Upstream
of collisionless shock, whistler waves are found to play a crucial
role in heating the plasma ions \cite{Scholer}. Their excitation and
propagation are not only limited to the Earth's nearby ionosphere, but
they are also found to be excited near the ionosphere of other planets
such as in the radiation belts of Jupiter and Saturn \cite{Bespalov}.
Whistlers are believed to be a promising candidate in transporting
fields and currents in plasma opening switch (POS) devices
\cite{mason}, which operate on fast electron time scales. Similarly,
electron whistlers have been known to drive the phenomenon of magnetic
field line reconnection \cite{bul} in astrophysical plasmas
\cite{zhou}.  Whistlers have also been investigated in several
laboratory experiments \cite{sten1,sten2,sten3,sten4}, where they have
been found to exhibit a variety of interesting features, such as an
anisotropic propagation of the phase front, strong dispersion
characteristics, interaction with plasma particles, etc.  A few
experimental features have also been confirmed by recent
three-dimensional simulations \cite{eliasson}, where it has been
reported that the polarity and the amplitude of the toroidal magnetic
field, in agreement with the laboratory experiments, determine the
propagation direction and speed of the whistlers.  These are only a
few examples amongst a large body of work devoted to the study of the
whistlers.  Despite the large amount of effort gone into
understanding the existence and the propagation of whistlers, their
linear dynamics is still debated, specially in the context of complex
nonlinear processes. For example, the role of whistlers in the high-
frequency turbulence and anisotropic spectral cascades has been
debated recently \cite{dastgeer1,dastgeer2,dastgeer3,dastgeer4,dastgeer5}.  
In the following, we present a summary of the spectra of the whistler
turbulence.

\begin{figure*}[ht]
\begin{center}
\includegraphics[width=14cm]{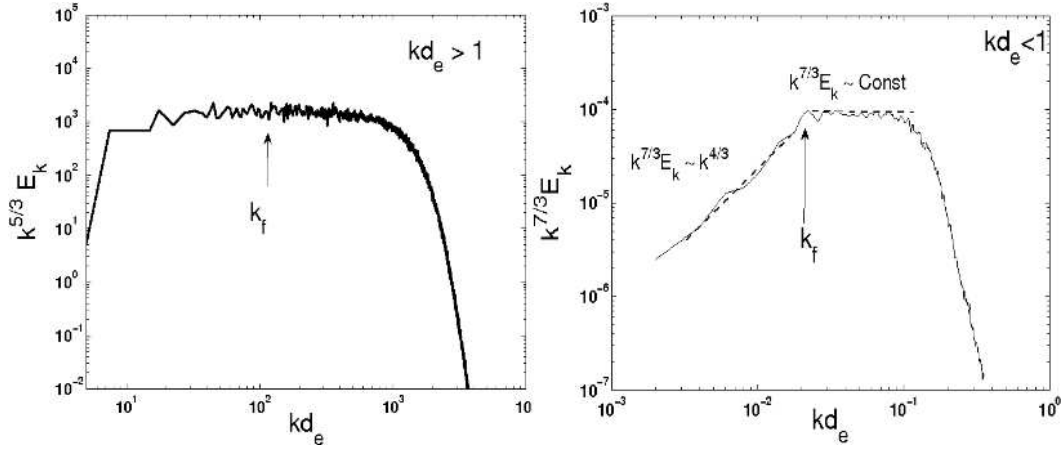}
\end{center}
\caption{ (left) Driven 2D whistler wave turbulence with the forcing band
identified by the $k_f$ modes.  The spectral resolution is $5120^2$.
Notice that no spectral break exists across the forcing modes, which
agrees with the Kolmogorov theory.  Across the $k_f$ modes, the
forward cascade of energy ($k>k_f$) and the inverse cascade of mean
magnetic potential ($k<k_f$) yield identical spectral indices and are
close to $E_k \sim k^{-5/3}$. We plot the compensated energy spectrum
$k^{5/3} E_k \sim \varepsilon^{2/3} \sim Const$ to show inertial range
modes along a straight line that is indicative of constant energy
dissipation rates in steady state driven dissipative turbulence.
(right) The inverse and forward cascade regimes are identified in 2D
driven whistler wave turbulence.  The compensated energy spectrum in
the inverse cascade regime $k^{7/3} E_{inv} \sim
\varepsilon^{2/3} k^{7/3} k^{-1} \sim \varepsilon^{2/3} k^{4/3}$,
whereas the forward cascade is given by $k^{7/3} E_{fwd} \sim
Const$. The approximate error in the spectral slope is about $\pm
0.072$. The spectral break, separating a forward and an inverse
cascade regimes, corresponds to a hump above the forcing modes $k_f$
that results from a direct excitation of the Fourier modes in driven
whistler wave turbulence.
}
\label{whistler}
\end{figure*}

The electron whistler mode dynamics is essentially governed by the 
EMHD equations. The EMHD phenomena typically occur on the electron time
scale [\cite{kingsep}], while the ions do not participate in the
whistler mode dynamics [\cite{shukla1}]. Thus, the basic frequency regimes involved are
$\omega_{ci} \ll \omega \ll \omega_{ce}$ (where $\omega_{ci},
\omega_{ce}$ are, respectively, the ion and electron gyrofrequencies, and
$\omega$ is the characteristic whistler frequency), and the length scales 
are $c/\omega_{pi} < \ell < c/\omega_{pe}$, where $\omega_{pi},
\omega_{pe}$ are the ion and electron plasma frequencies, respectively. 
In whistlers, electrons carry currents, since immobile ions 
merely provide a neutralizing background to a quasi-neutral EMHD plasma.
The electron momentum equation is
\be
\label{elec}
m_e n \frac{d}{dt}{\bf V}_e
=-en {\bf E} -  \frac{ne}{c} {\bf V}_e \times {\bf B} - \nabla P
-\mu m_e n {\bf V}_e, 
\ee
\be
{\bf E} = -\nabla \phi - \frac{1}{c} \frac{\partial {\bf A}}{\partial t},
\ee
\be
\label{ampere}
\nabla \times {\bf B} = \frac{4\pi}{c} {\bf J} +  
\frac{1}{c} \frac{\partial {\bf E}}{\partial t},
\ee
\be
\frac{\partial n}{\partial t} + \nabla \cdot (n {\bf V}_e) = 0,
\ee
where \[\frac{d}{dt}=\frac{\partial}{\partial t} + {\bf V}_e \cdot \nabla.  \]
The remaining equations are ${\bf B} = \nabla \times {\bf A}, {\bf J}
= -en{\bf V}_e, \nabla \cdot {\bf B} =0$. Here $m_e, n, {\bf V}_e$ are
the electron mass, density and fluid velocity respectively. ${\bf E},
{\bf B}$ respectively represent electric and magnetic fields and $\phi
, {\bf A}$ are electrostatic and electromagnetic potentials. The
remaining variables and constants are, the pressure $P$, the
collisional dissipation $\mu$, the current due to electrons flow ${\bf
J}$, and the velocity of light $c$.  The displacement current in
Amp\'ere's law \eq{ampere} is ignored, and the density is considered as
constant throughout the analysis.  The electron continuity equation
can, therefore, be represented by a divergence-less electron fluid
velocity $\nabla \cdot {\bf V}_e = 0$. The electron fluid velocity can
then be associated with the rotational magnetic field through
\[
{\bf V}_e = - \frac{c}{4\pi n e}\nabla \times{\bf B}.
\]

By taking the  curl of \eq{elec} and, after slight rearrangement
of the terms, we obtain a generalized electron momentum equation
in the following form.
\be
\label{PP}
\frac{\partial {\bf P}}{\partial t} - {\bf V}_e  \times ( \nabla \times {\bf P})
+\nabla \xi = - \mu m_e {\bf V}_e
\ee
where
\[
{\bf P} = m_e{\bf V}_e - \frac{e{\bf A}}{c} ~~~~~~~{\rm and}~~~
\xi = \frac{1}{2} m_e {\bf V}_e\cdot {\bf V}_e + \frac{P}{n} - e\phi.
\]

The curl of \eq{PP} leads to 
a three-dimensional equation of EMHD describing the evolution of
the whistler wave magnetic field, 
\be
\label{emhd3}
\frac{\partial }{\partial t}{\bf \Omega}_B  + {\bf
V}_e\cdot \nabla {\bf \Omega}_B - {\bf \Omega}_B 
 \cdot \nabla{\bf V}_e= \mu d_e^2 \nabla^2 {\bf B}.
\ee 
where ${\bf \Omega}_B={\bf B} - d_e^2\nabla^2 {\bf B}, d_e=c/\omega_{pe}$, the electron skin depth, 
is an intrinsic length-scale in the EMHD plasma.  The three-dimensional EMHD equations can be
transformed into two dimensions by regarding variation in the
$\hat{z}$-direction as ignorable i.e.  $\partial/\partial z=0$, and
separating the total magnetic field ${\bf B}$ into two scalar
variables, such that ${\bf B} = \hat{z} \times \nabla \psi +
b\hat{z}$.  Here $\psi$ and $b$ respectively present perpendicular and
parallel components of the wave magnetic field. The corresponding
equations of these components can be written in a normalized form as follows,
\be
\label{psi}
 \frac{\partial}{\partial t}\Omega_\psi +
{\hat{z}\times{\bf \nabla}b} \cdot \nabla \Omega_\psi 
- B_0\frac{\partial}{\partial y}b
=0,
\ee 
\be 
\label{b}
\frac{\partial}{\partial t} \Omega_B -
d_e^2{\hat{z}\times{\bf \nabla}b} \cdot \nabla \nabla^2 b +
{\hat{z}\times{\bf \nabla}\psi} \cdot \nabla \nabla^2 \psi 
+ B_0 \frac{\partial}{\partial y}\nabla^2 \psi 
=0,
\ee
where $\Omega_\psi=\psi - d_e^2\nabla^2 \psi, \Omega_B=b -
d_e^2\nabla^2 b$ The length and time scales are normalized
respectively by $d_e$ and $\omega_{ce}$, whereas magnetic field is
normalized by a typical mean $B_0$. The linearization of \eqs{psi}{b}
about a constant magnetic field $B_0$ yields the dispersion relation
for the whistlers, the normal mode of oscillation in the EMHD
frequency regime, and is given by
\[
\omega_k = \omega_{c_0}\frac{d_e^2 k_yk}{1+d_e^2k^2},
\]
where $ \omega_{c_0}=eB_0/mc$ and $k^2=k_x^2+k_y^2$.  From the set of
 the EMHD \eqs{psi}{b}, there exists an intrinsic length scale $d_e$,
 the electron inertial skin depth, which divides the entire spectrum
 into two regions; namely short scale ($kd_e>1$) and long scale
 ($kd_e<1$) regimes.  In the regime $kd_e<1$, the linear frequency of
 whistlers is $\omega_k \sim k_y k$ and the waves are dispersive.
 Conversely, dispersion is weak in the other regime $kd_e>1$ since
 $\omega_k \sim k_y/ k$ and hence the whistler wave packets interact
 more like the eddies of hydrodynamical fluids.

\section{Energy spectra in whistler turbulence}

By virtue of $d_e$, there exists two inertial ranges that correspond
to smaller $kd_e>1$ and larger $kd_e<1$ length-scales in the EMHD
 turbulence. Correspondingly, forward cascade turbulent spectra in
 these regimes exhibit $k^{-5/3}$ and $k^{-7/3}$ respectively as shown
 in \Fig{whistler}. The whistler spectrum $k^{-7/3}$ is produced
 essentially by fluctuations in electron fluid while {\it ions are at
   rest}.  The observed spectra in our simulations depicted as in
 \Fig{whistler} can be explained on the basis of Kolmogorov
 phenomenology as follows.

The regime $kd_e > 1$ in EMHD turbulence corresponds essentially to a
hydrodynamic regime because the energy spectrum is dominated by the
shorter length-scale turbulent eddies that give rise to a
characteristic spectrum of an incompressible hydrodynamic fluid. The
group velocity of whistlers, in this regime, is small and hence it is
expected that the whistler effect cannot be present.  For $kd_e > 1$,
the first term in the magnetic potential can be dropped and thus
\[ A \sim (\nabla^2 \psi)^2. \]
Similarly the dominating terms in the energy are
\[ E \sim (\nabla b)^2 + (\nabla^2 \psi)^2. \]
  This implies 
\[ E \sim (\nabla b)^2 \sim (\nabla^2 \psi)^2. \]
Thus for $kd_e > 1$, $E \sim A$.  In this case both magnetic potential
and the energy cascades turn out to be identical leading to the same
spectral index, i.e. $k^{-5/3}$, in both direct and inverse cascade
regimes. This is shown in \Fig{whistler} (left panel).

In EMHD the eddy velocity ($v$) in the $x-y$ plane is characterized by $
{\hat{z}\times{\bf \nabla }b}$. Thus the typical velocity with a scale
size $\ell$ can be represented by $v \simeq b_{\ell}/\ell$. The eddy
scrambling time is then given by
\[\tau \sim \frac{\ell}{v} \sim \frac{\ell^2}{b_{\ell}}.\]
In the limit of $kd_e < 1$, the mean square magnetic potential is,
\[ A \sim \psi^2 \sim b_{\ell}^2 \ell^2. \]
The second similarity follows from the assumption of an equipartition
of energy in the axial and poloidal components of magnetic field. This
assumption has been invoked in some earlier works too, and would be
exact if the EMHD turbulence comprised of randomly interacting
whistler waves only.  To determine the scaling behaviour of the
spectrum in the magnetic potential cascade regime (i.e. for $k < k_f$,
$k_f$ representing the forcing scales), the rate of $A$ transfer is
given by the relation,
\[ \varepsilon = \frac{A}{\tau} = b_{\ell}^3, \]
and the mean square magnetic potential per unit wave number is
\[ A_k = b_{\ell}^2 \ell^3 .\]
The locality of the spectral cascade in the wavenumber space yields $
A_k = \varepsilon^{\alpha}k^{\beta}$. On equating the powers of $b_{\ell}$
and $\ell$, we obtain $\alpha = 2/3$ and $\beta = -3$. In $kd_e < 1$
regime, the expression for the energy can be approximated as
$ E \sim b^2 \sim (\nabla \psi)^2$. This further 
 implies that $E = A/\ell^2$ and 
\[ E_k \sim k^{-1}. \]
 A similar analysis in the other regime of $k > k_f$ where energy
 cascade is local in wave number space leads to an energy spectrum
 with the index of $-7/3$ such that
\[ E_k \sim k^{-7/3}. \]
The compensated spectra of \Fig{whistler} (right panel) are consistent with 
above description.

 In the following section, we shall include the ion dynamics that
 correspond to turbulence scales in which $kd_i>1$, where
 $d_i$ is ion skin depth. This regime of the magnetofluid
 dynamics is referred to as the Hall MHD, and it stems from the combined motion
 of the electron and ion fluids amidst predominant density fluctuations
 that evolves on Kinetic-Alfv\'en wave (KAW) time (and length) scales.

\section{The High-Frequency Kinetic-Alfv\'en Inertial Regime}

The high-frequency kinetic-Alfv\'en regime of the plasma results in
the solar wind turbulence when plasma wave fluctuations comprise a
multitude of time and space scales that are shorter than the ion
gyroperiod ($2\pi/\Omega_{ci}$) and comparable (or longer) to (than)
the ion thermal gyroradius/ion-sound gyroradius or shorter than the
ion skin depth, respectively. These electromagnetic fluctuations
cannot be described by the usual magnetohydrodynamic (MHD) model
\citep{danskat,goldstein1,leamon,goldstein2,dastgeer7,dastgeer8} which
involve the characteristic time scale much longer than the ion
gyroperiod.  Typically, reconnection of magnetic field lines near the
solar corona \citep{amitava1}, Earth's magnetosphere and in many
laboratory plasma devices \citep{amitava2} are other avenues where
collisionless or kinetic processes call for inclusion of the finite
Larmor radius and finite frequency effects in the standard MHD
model. In particular, in the context of the solar wind plasma, higher
time resolution databases identify a spectral break near the end of
the MHD inertial range spectrum that corresponds to a high-frequency
($>\Omega_{ci}$) regime where turbulent cascades are {\em not}
explainable by Alfv\'enic cascades.  This refers to a secondary
inertial range where turbulent cascades follow a $k^{-7/3}$ (where $k$
is a typical wavenumber) spectrum in which the characteristic
fluctuations evolve typically on kinetic-Alfv\'en time scales
\citep{howes08,matthaeus08}. The onset of the second or the
kinetic-Alfv\'en inertial range still eludes our understanding of
solar wind fluctuations \citep{leamon,dastgeer7,dastgeer8}. The
mechanism leading to the spectral break has been thought to be either
mediated by the kinetic- Alfv\'en waves (KAWs), or damping of
ion-cyclotron waves, or dispersive processes, or other implicit
nonlinear interactions associated with kinetic effects in solar wind
plasmas \citep{danskat,goldstein1,leamon,goldstein2,dastgeer7,dastgeer8}.
Motivated by these issues, we have developed three dimensional, time
dependent, compressible, non-adiabatic, driven and fully parallelized
Hall-magnetohydrodynamic (H-MHD) simulations
\citep{dastgeer7,dastgeer8} to investigate turbulent spectral cascades
in collisionless space plasmas.

In the high-frequency regime, $\omega > \Omega_{ci}$, the inertialess
electrons contribute to the electric field which is dominated
essentially by the Hall term corresponding to ${\bf J}
\times {\bf B}$ force.  The latter, upon substituting in the ion
momentum equation, modifies the ion momentum, the magnetic field and total
energy in a manner to introduce a high-frequency ($\omega >
\Omega_{ci}$) and small scale ($k_\perp \rho_L>1$, where $\rho_L$ is
ion thermal gyroradius) plasma motions. The characteristic length scales
($k^{-1}$) associated with the plasma motions are smaller than the ion
gyroradius ($\rho_L$).  The quasi-neutral solar wind plasma density
($\rho$), velocity (${\bf U}$), magnetic field (${\bf B}$) and total
pressure ($P=P_e+P_i$) fluctuations can then be cast into a set of
the Hall-MHD equations, given by
\be
\label{mhd:cont}
\frac{\partial \rho}{\partial t} + \nabla \cdot (\rho{\bf U})=0,
\ee
\be
\label{mhd:mom}
\rho \left(  \frac{\partial }{\partial t} + {\bf U} \cdot \nabla \right) {\bf U}
= -\nabla P + \frac{1}{c} {\bf J} \times {\bf B}
\ee
\be
\label{mhd:mag}
 \frac{\partial {\bf B}}{\partial t} = \nabla \times \left({\bf U} \times {\bf B}-
d_i \frac{{\bf J} \times {\bf B}}{\rho} \right)+\eta \nabla^2 {\bf B},
\ee
\be
\label{mhd:en}
 \frac{\partial e}{\partial t} + \nabla \cdot \left( \frac{1}{2}\rho
 U^2{\bf U} + \frac{\gamma}{\gamma-1}\frac{P}{\rho}\rho{\bf U} +
 \frac{c}{4\pi}{\bf E} \times {\bf B} \right) =0 \ee 
where 
\[ e=\frac{1}{2}\rho U^2 + \frac{P}{(\gamma-1)}+\frac{B^2}{8\pi}\]
 is the total energy of the 
plasma that contains both the electron and ion motions.  All the dynamical
variables are functions of three space and a time, i.e. $(x,y,z,t)$,
co-ordinates. Equations (\ref{mhd:cont}) to (\ref{mhd:en}) are
normalized by typical length $\ell_0$ and time $t_0 = \ell_0/v_0$
scales in our simulations, where $v_0=B_0/(4\pi \rho_0)^{1/2}$ is
Alfv\'en velocity such that $\bar{\nabla}=\ell_0{\nabla},
\partial/\partial \bar{t}=t_0\partial/\partial t, \bar{\bf U}={\bf
  U}/v_0,\bar{\bf B}={\bf B}/v_0(4\pi \rho_0)^{1/2},
\bar{P}=P/\rho_0v_0^2, \bar{\rho}=\rho/\rho_0$.  The parameters $\mu$
and $\eta$ correspond respectively to ion-electron viscous drag term
and magnetic field diffusivity. While the viscous drag modifies the
dissipation in plasma momentum in a nonlinear manner, the magnetic
diffusion damps the small scale magnetic field fluctuations linearly.
The magnetic field is measured in the unit of Alfv\'en velocity. The
dimensionless parameter in magnetic field \eq{mhd:mag} i.e. ion skin
depth $\bar{d}_i=d_i/\ell_0, d_i=C/\omega_{pi}$ is associated with the
Hall term.  This means the ion inertial scale length ($d_i$) is a
natural or an intrinsic length scale present in the Hall MHD model
which accounts for finite Larmour radius effects corresponding to high
frequency oscillations in $kd_i>1$ regime.  Clearly, the Hall force
dominates the magnetoplasma dynamics when $1/\rho({\bf J} \times {\bf
  B}) > {\bf U} \times {\bf B}$ term in \eq{mhd:mag} which
in turn introduces time scales corresponding to the high frequency
plasma fluctuations in $kd_i>1$ regime.  Furthermore, our model
includes a full energy equation [\eq{mhd:en}] unlike an adiabatic
relation between the pressure and density. The use of energy equation
enables us to study a self-consistent evolution of turbulent heating
resulting from nonlinear energy cascades in the solar wind plasma.

\begin{figure}[t]
\includegraphics[width=8cm]{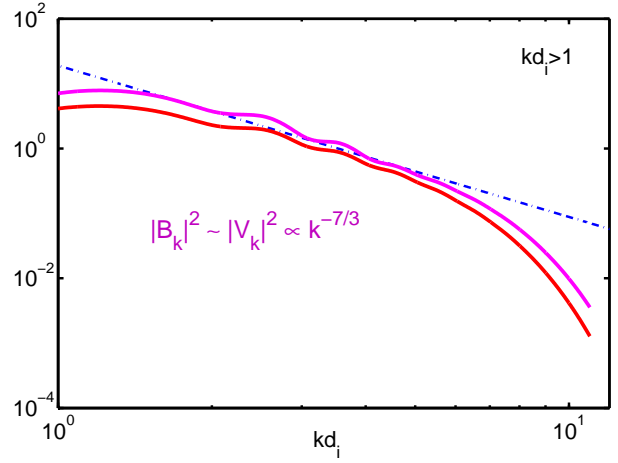}
\caption{Inertial range turbulent spectra of magnetic and velocity
  field fluctuations are shown. These fluctuations closely follow a
  $k^{-7/3}$ spectrum in kinetic Alfv\'en regime where $kd_i>1$.  The
  simulation parameters are: spectral resolution is $128^3$,
  $\eta=\mu=10^{-3}, M_A=\beta=1.0, kd_i \ge 1, L_x=L_y=L_z=2\pi$.}
\label{spectra}
\end{figure}

\begin{figure}[t]
\includegraphics[width=9cm]{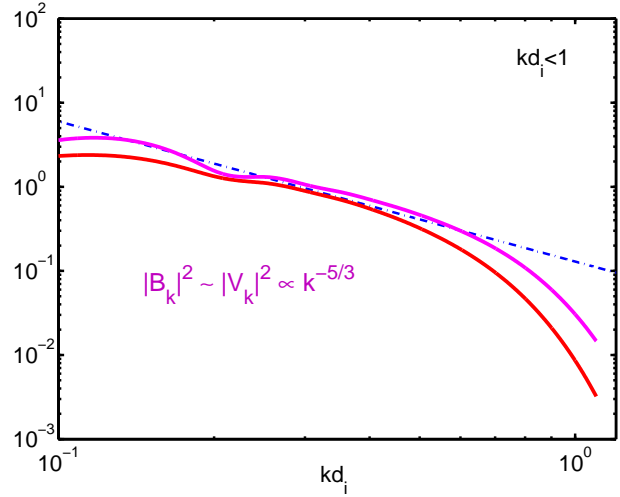}
\caption{Inertial range turbulent spectra in $kd_i<1$ regime. Shown
  are the magnetic and velocity field fluctuations along with that of
  the density fluctuation. These fluctuations closely follow a
  $k^{-5/3}$ omnidirectional spectrum in the Alfv\'enic
  regime. Simulation parameters are the same as in \Fig{spectra},
  except $d_i=0.05$.}
\label{spectra2}
\end{figure}

In the linear regime, the Hall-MHD equations (without the energy equation) 
admit a dispersion relation, 
\[(\omega^ 2-k_z^ 2V_A^ 2)D_m (\omega, {\bf k})
=\omega^ 2(\omega^ 2-k^ 2V_s^ 2) k_z^ 2k ^2V_A^ 4/\omega_{ci}^ 2,\]
which exhibits coupling among the fast and slow magnetosonic waves
given by the solutions of 
\[D_m(\omega, {\bf k}) =\omega^ 4 -\omega^ 2
k^ 2 (V_A^ 2+V_s^ 2)+k_z^ 2k^ 2 V_A^ 2 V_s^ 2=0,\] where $\omega$ is
the frequency, ${\bf k} (={\bf k}_\perp + k_z \hat {\bf z}$ is the
wave vector, the subscripts $\perp$ and $z$ represent the components
across and along the external magnetic fields $B_0 \hat {\bf z}$, and
$V_A$ and $V_s$ are the Alfv\'en and effective sound speeds,
respectively. The warm HMHD plasma thus supports a great variety of
waves (e.g. magnetic field-aligned non-dispersive Alfv\'en waves,
obliquely propagating fast and slow magnetohydrodynamic waves,
obliquely propagating dispersive kinetic Alfv\'en waves and whistlers)
having different wavelengths (comparable to the ion skin depth
associated with the Hall drift, the ion sound gyroradius associated
with the electron pressure and the perpendicular ion inertia/ion
polarization drift). Interestingly, in the $\omega < \Omega_{ci}$
regime, which is predominantly known to be an Alfv\'enic regime,
\citet{howes08} noted the possibility of occurance of highly oliquely
propagating KAWs (with $\omega \ll \Omega_{ci}$) thereby questioning
the role of damping of ion cyclotron waves in the onset of spectral
breakpoint. While our simulations described in \Fig{spectra} may
contain the KAWs corresponding to $\omega \ll \Omega_{ci}$, their
specific contribution in the spectral cascades depicted in
\Fig{spectra} and \Fig{spectra2} is not identifiable. We therefore
leave this study to a future investigation.

The nonlinear spectral cascades in the KAW regime lead to a secondary
inertial range in the vicinity of modes $kd_i \simeq 1$ in that the
solar wind magnetic, velocity and field fluctuations follow turbulent
spectra close to $k^{-7/3}$. This is shown in \Fig{spectra}. The
characteristic turbulent spectra in the KAW regime is steeper than
that of MHD inertial range which follows a $k^{-\alpha}$ where
$\alpha$ is $5/3$ or $3/2$. The onset of the secondary inertial range
is constantly debated because of the presence of multiple processes in
the KAW regime that include, for instance, the dispersion, damping of
ion cyclotron waves, turbulent dissipation or due to fast or slow
magnetosonic perturbations and etc. In the context of our simulations, the
observed $k^{-7/3}$ in the solar wind turbulent plasma can be
understood from the energy cascades affected by the Hall forces. The
latter are one of the potential candidates that may be responsible for
a $k^{-7/3}$ spectrum in the KAW inertial range regime.

To understand the observed $k^{-7/3}$ spectrum in the KAW inertial
range regime in solar wind plasma and in our simulations, we invoke
Kolmogorov and Kraichnan like phenomenologies \citep{r17,krai} in Hall
MHD plasma.  The time scale associated with Hall MHD is $\tau_{\rm H}$
smaller than that of MHD $\tau_{\rm MHD}$.  The nonlinear cascades in
the MHD turbulence are governed typically by $\tau_{\rm MHD} \sim (k
u_k)^{-1}$, where $u_k$ is the velocity field in the $k$-space. By
contrast, the spectral transfer of the turbulent energy in the solar
wind (Hall MHD) plasma is $\tau_{\rm H} \sim (k^2 B_k)^{-1}$. The
energy transfer rates in the KAW regime are therefore $\varepsilon
\sim u_k^2/\tau_{\rm H}$. On substituting the turbulent equipartition
relation between the the velocity and magnetic fields $B_k^2 \sim
u_k^2$, the energy transfer rates turn out to be $\varepsilon \sim k^2
B_k^3$.  The use of the turbulent equipartition is justified from the
observations and it is also seen in our 3D simulations [see
  Fig. (5)]. Applying the Kolmogorov's phenomenology \cite{r17,krai}
that the energy cascades in the inertial range are local and that they
depend on the Fourier modes and the energy dissipation rate, we obtain
$k^{-1}B_k^2 \sim (B_k^3 k^2)^{\alpha} k^{\beta}$. Equating the
indices of the common bases, we get $\alpha = 2/3$ and $\beta =
-7/3$. This results in an energy spectrum $E_k \propto k^{-7/3}$ that
is consistent with our 3D simulations [see \Fig{spectra}] and the
observations. On the other hand, the use of $\tau_{\rm MHD}$ in
estimating the energy dissipation rates retrieves the inertial range
MHD spectrum. This indicates that the Hall effects may be responsible
for the spectral steepening in the solar wind plasma fluctuations in
the $kd_i>1$ regime.  It is worth noted that our simulations in
$kd_i<1$ exhibit MHD like $k^{-5/3}$ omnidirectional Kolmogorov-like
\cite{r17} spectrum as shown in \Fig{spectra2}.

\conclusions

In summary, we have presented a survey of the spectral properties of
low-frequency electromagnetic turbulence in uniform magnetized plasmas. 
Specifically, we have focused on the electron whistlers, and high-frequency 
kinetic Alfv\'en waves. While for the electron whistlers the motion of ions 
is unimportant, the ion dynamics plays an essential role in the
high-frequency KAW turbulence.  We have developed codes to numerically solve
the governing nonlinear equations for the whistler mode and KAW turbulences.
Our simulations for the whistler mode turbulence depict a dual cascade process 
between the energy and mean magnetic potential in consistent with a Kolmogorov
phenomenology.  Furthermore, the high-frequency kinetic Alfv\'en
regime, typically observed in the solar wind plasma, is studied by
means of 3D simulations. Our 3D simulation results exhibit that the
characteristic turbulent interactions in the high-frequency regime
evolve typically on the Hall time-scales and modify the inertial range
spectra that eventually leads to the observed spectral break in the
solar wind.


The support of NASA(NNG-05GH38) and NSF (ATM-0317509) grants is
acknowledged.

This work was partially supported by the Deutsche
Forschungsgemeinschaft through the project SH 21/3-1 of the 
Forschergruppe 1048.












\addtocounter{figure}{-1}\renewcommand{\thefigure}{\arabic{figure}a}


\begin{thebibliography}{}



\bibitem[Bale et al. (2005)]{bale05} Bale, S. D., Kellogg, P. J.,
  Mozer, F. S., Horbury, T. S., and Reme, H., Measurement of the
  Electric Fluctuation Spectrum of Magnetohydrodynamic Turbulence.
  Phys. Rev. Lett. 94, 215002 (2005).

\bibitem[Bespalov (2006)]{Bespalov}
{Bespalov, P. A.}, Excitation of whistler waves in three spectral
bands in the radiation belts of Jupiter and Saturn.  {\em European
Planetary Science Congress}, Berlin, Germany, 18 - 22 September 2006.,
p.461 (2006).


\bibitem[Biskamp et al. (1996)]{biskamp96}
Biskamp, D., Schwarz, E., and 
Drake, J. F., Two-dimensional electron magnetohydrodynamic turbulence. Phy. Rev. Lett., {\bf 76}  1264, 1996



\bibitem[Biskamp \& Bremer (1994)]{biskamp93} Biskamp, D. \& Bremer,
  U., Dynamics and Statistics of Inverse Cascade Processes in 2D
  Magnetohydrodynamic Turbulence. Phys. Rev. Lett. 72, 3819 (1994).





\bibitem[Bhattacharjee (2004)]{amitava1}
Bhattacharjee, A., Impulsive Magnetic Reconnection in the Earth's Magnetotail and the Solar Corona.
Annu. Rev. Astron. Astrophys {\bf 42}, 365 (2004).

\bibitem[Bhattacharjee et al (2001)]{amitava2}
Bhattacharjee, A.; Ma, Z. W.; Wang, Xiaogang.,
Recent developments in collisionless reconnection theory: Applications to laboratory and space plasmas.
Phys. Plasmas {\bf 8}, 1829 (2001).


\bibitem[Bulanov et al (1992)]{bul}
{Bulanov, S. V., Pegoraro, F.,  and  Sakharov, A. S.}, 
Magnetic reconnection in electron magnetohydrodynamics.
{\it Phys. Fluids}, B {\bf 4} 2499 (1992).

\bibitem[Cattell et al (2008)]{cattell}	
{Cattell, C.; Wygant, J. R.; Goetz, K.; Kersten, K.; Kellogg, P. J.;
von Rosenvinge, T.; Bale, S. D.; Roth, I.; Temerin, M.; Hudson, M. K.;
Mewaldt, R. A.; Wiedenbeck, M.; Maksimovic, M.; Ergun, R.; Acuna, M.;
Russell, C. T.},
Discovery of very large amplitude whistler-mode waves in Earth's radiation belts.
{\em Geophys. Res. Lett.}, {\bf 35},  L01105 (2008). 

\bibitem[Denskat et al (1983)]{danskat}
Denskat, K. U.; Beinroth, H. J.; Neubauer, F. M.
Interplanetary magnetic field power spectra with frequencies from 2.4
X 10 to the -5th HZ to 470 HZ from HELIOS-observations during solar
minimum conditions.
J. Geophys. Res. {\bf 54}, 60 (1983).

\bibitem[Eliasson \& Shukla (2007)]{eliasson}
{Eliasson, B.; Shukla, P. K.},
Dynamics of Whistler Spheromaks in Magnetized Plasmas.
{\em Phys. Rev. Lett.} {\bf 99}, 205005 (2007).


\bibitem[Frisch (1995)]{fri}
Frisch, U., Turbulence: The Legacy of A.N. Kolmogorov 
(Cambridge University Press, Cambridge, 1995).



\bibitem[Fyfe \& Montgomery (1977)]{fyfe} Fyfe, D., Montgomery, D.,
  and Joyce, G., Dissipative, forced turbulence in two-dimensional
  magnetohydrodynamics. J. Plasma Phys. 17, 369 (1977).

\bibitem[Goldstein et al (1994)]{goldstein1}
Goldstein, M. L.; Roberts, D. A.; Fitch, C. A.
Properties of the fluctuating magnetic helicity in the inertial and dissipation ranges of solar wind turbulence.
J. Geophys. Res. {\bf 99}, 11519 (1994).

\bibitem[Goldstein et al (1996)]{goldstein2}
Ghosh, S.; Siregar, E.; Roberts, D. A.; Goldstein, M. L.
Simulation of high-frequency solar wind power spectra using Hall magnetohydrodynamics.
J. Geophys. Res. {\bf 101}, 2493 (1996).


\bibitem[Howes et al. (2008)]{howes08} Howes, G. G., Dorland, W.,
  Cowley, S. C., Hammett, G. W., Quataert, E., Schekochihin, A. A.,
  and Tatsuno, T., Kinetic Simulations of Magnetized Turbulence in
  Astrophysical Plasmas. Phys. Rev. Lett. 100, 065004 (2008).

\bibitem[Helliwell (1965)]{helli}
{Helliwell, A.}, 
Whistlers and Related Ionospheric Phenomena.
Stanford University Press, Stanford, CA. (1965).


\bibitem[Kraichnan (1965)]{krai}
Kraichnan, R. H., Inertial-Range Spectrum of Hydromagnetic Turbulence.
Phys. Fluids {\bf 8}, 1385 (1965).


\bibitem[Kingsep et al (1990)]{kingsep}
{Kingsep, A. S., Chukbar, K. V., and  Yan'kov V. V.}
Reviews of Plasma
Physics. Consultants Bureau, New York, Vol {\bf 16} (1990).\\
Gordeev, A. V., Kingsep, A. S., and Rudakov, L. I.,  
{\em Phys. Reports}, {\bf 243}, 215--315 (1994).

\bibitem[Shukla (1978)]{shukla1}
{Shukla, P. K.},
Modulational instability of whistler-mode signals
{\em Nature} {\bf 274}, 874 (1978).


\bibitem[Kolmogorov (1951)]{r17}
 Kolmogorov, A. N. 1951 
The Local Structure of Turbulence in Incompressible Viscous Fluid for Very Large Reynolds' Numbers.
{\it C. R. Acad. Sci. U. R. S. S.} {\bf 30}, 301,
and {\bf 30}, 538 (1941).

\bibitem[Leamon et al (1998)]{leamon}
Leamon, Robert J.; Matthaeus, William H.; Smith, Charles W.; Wong, Hung K.
Contribution of Cyclotron-resonant Damping to Kinetic Dissipation of Interplanetary Turbulence.
Astrophys. J. {\bf 507}, L181 (1998).


\bibitem[Matthaeus et al. (2008)]{matthaeus08} Matthaeus, W. H.,
  Servidio, S., and Dmitruk, P., Comment on “Kinetic Simulations of
  Magnetized Turbulence in Astrophysical Plasmas”.
  Phys. Rev. Lett. 101, 149501 (2008).

\bibitem[Mason et al (1993)]{mason}	
{Mason, R. J., Auer, P. L., Sudan, R. N.,  Oliver, B. E.,  Seyler, C. E.,  and
Greenly, J. B.}, 
Nonlinear magnetic field transport in opening switch plasmas.
{\it Phys Fluids} B {\bf 5}  1115 (1993).



\bibitem[Russell et al (2007)]{Russell}
{Russell, C. T.; Zhang, T. L.; Delva, M.; Magnes, W.; Strangeway, R. J.; Wei, H. Y.},
Lightning on Venus inferred from whistler-mode waves in the ionosphere.
{\em Nature}, {\bf 450}, Issue 7170,  661-662 (2007). 


\bibitem[Servidio et al. (2008)]{servidio08} Servidio, S., Matthaeus,
  W. H., and Carbone, V., Statistical properties of ideal
  three-dimensional Hall magnetohydrodynamics: The spectral structure
  of the equilibrium ensemble. Phys. Plasmas 15, 042314 (2008).

\bibitem[Scholer \&  Burgess (2007)]{Scholer}
{Scholer, M., and  Burgess, D.},
Whistler waves, core ion heating, and non-stationarity in oblique collisionless shocks.
{\em Phys. Plasmas}, {\bf 14}, 072103-072103-11 (2007).

\bibitem[Stenberg et al (2007)]{Stenberg}	
{Stenberg, G.; Oscarsson, T.; André, M.; Vaivads, A.; Backrud-Ivgren,
M.; Khotyaintsev, Y.; Rosenqvist, L.; Sahraoui, F.;
Cornilleau-Wehrlin, N.; Fazakerley, A.; Lundin, R.; Décréau, P. M. E.},
Internal structure and spatial dimensions of whistler wave regions in the magnetopause boundary layer.
{\em Annales Geophysicae}, {\bf 25}, 11, 2439-2451 (2007).


\bibitem[Stenzel (1997)]{sten1} 
{Stenzel, R. L.}, 
Whistler wave propagation in a large magnetoplasma.
{\it Phys. Fluids}, {\bf 19}, No. 6, 857 (1976).

\bibitem[Stenzel (1975)]{sten2} 
{Stenzel, R. L.}, 
Self-ducting of large-amplitude whistler waves.
{\it Phy. Rev. Lett.}, {\bf 35}, No. 9, 574  (1975).


\bibitem[Stenzel \& Urrutia (1990)]{sten3} 
{Stenzel, R. L.,  Urrutia, J. M.},
Force-free electromagnetic pulses in a laboratory plasma.
{\it Phy. Rev. Lett.}, {\bf 65}, No. 16, 2011 (1990).


\bibitem[Stenzel et al (1993,19940)]{sten4} 
{Stenzel, R. L.,  Urrutia, J. M., and Rousculp, C. L.},
Pulsed currents carried by whistlers. I - Excitation by magnetic antennas.
{\it Phys. Fluids}, B {\bf 5}, No. 2, 325 (1993);
{Urrutia, J. M.,  Stenzel, R. L.,  and Rousculp, C. L.},
Pulsed currents carried by whistlers. II. Excitation by biased electrodes.
{\it Phys. Plasmas},  {\bf 1}, No. 5,  1432 (1994).


\bibitem[Shaikh et al (2000a)]{dastgeer1}
{Shaikh, D.,  Das, A., Kaw, P. K., Diamond, P.},
Whistlerization and anisotropy in two-dimensional electron magnetohydrodynamic turbulence.
{\em Phys. Plasmas} {\bf 7}, 571 (2000).

\bibitem[Shaikh et al (2000b)]{dastgeer2}
{Shaikh, D., Das, A., and Kaw, P. K.},
Hydrodynamic regime of two-dimensional electron magnetohydrodynamics. 
{\em  Phys. Plasmas} {\bf 7}, 1366 (2000).


\bibitem[Shaikh \& Zank (2003)]{dastgeer3}
{Shaikh, D.,  and Zank, G. P.},
Anisotropic Turbulence in Two-dimensional Electron Magnetohydrodynamics. 
{\em Astrophys. J.} {\bf 599}, 715 (2003).

\bibitem[Shaikh (2004)]{dastgeer4}
{Shaikh, D.},
Generation of Coherent Structures in Electron Magnetohydrodynamics.
{\em Physica Scripta}, {\bf 69}, 216 (2004).


\bibitem[Shaikh  (2008)]{dastgeer6}
Shaikh, D.,
Theory and Simulations of Whistler Wave Propagation
{\em  J. Plasma Phys.} {\bf 75}, 117 (2008).

\bibitem[Shaikh (2009)]{dastgeer2009} Shaikh, D., Whistler Wave
  Cascades in Solar Wind Plasma, {\em Mon Not Royal Astrono. Soc.}, In
  press, (2009).

\bibitem[Shaikh \& Zank (2005)]{dastgeer5}
{Shaikh, D.,  and Zank, G. P.},
Driven dissipative whistler wave turbulence. 
{\em  Phys. Plasmas} {\bf 12}, 122310 (2005).


\bibitem[Shaikh \& Shukla (2008)]{dastgeer7} 
{Shaikh, D. and Shukla,  P. K.}, 
Three Dimensional Simulations of Compressible Hall MHD
  Plasmas.  {\em AIPC}, FRONTIERS IN MODERN PLASMA PHYSICS: 2008 ICTP
  International Workshop on the Frontiers of Modern Plasma
  Physics. AIP Conference Proceedings, Volume 1061, pp. 66-75.

\bibitem[Shaikh \& Shukla (2009)]{dastgeer8}
{Shaikh, D. and Shukla,  P. K.}, 
3D simulations of fluctuation spectra in the Hall-MHD plasma.
{\em Phys. Rev. Lett.},   102, 045004 (2009)


\bibitem[Wei et al (2007)]{wei}	
{Wei, X. H.; Cao, J. B.; Zhou, G. C.; Santolík, O.; Rème, H.;
Dandouras, I.; Cornilleau-Wehrlin, N.; Lucek, E.; Carr, C. M.;
Fazakerley, A.}, Cluster observations of waves in the whistler
frequency range associated with magnetic reconnection in the Earth's
magnetotail.
{\em Journal of Geophysical Research}, {\bf 112}, A10, A10225 (2007). 


\bibitem[Zhou et al (1996)]{zhou} {Zhou, H. B., Popadopolous, K.,
  Sharma A. S., and Chang, C. L.}, Electronmagnetohydrodynamic
  response of a plasma to an external current pulse.  {\it
    Phys. Plasmas}, {\bf 3} 1484 (1996).


\end{thebibliography}
\end{document}